\documentclass[10pt]{article}
\usepackage[utf8]{inputenc}
\usepackage[T1]{fontenc}
\usepackage{amsmath}
\usepackage{amsfonts}
\usepackage{amssymb}
\usepackage[version=4]{mhchem}
\usepackage{stmaryrd}
\usepackage{hyperref}
\hypersetup{colorlinks=true, linkcolor=blue, filecolor=magenta, urlcolor=cyan,}
\usepackage{graphicx}
\usepackage[export]{adjustbox}
\graphicspath{ {./images/} }

\title{SLENet: A Novel Multiscale CNN-Based Network for Detecting the Rats Estrous Cycle }

\author{Qinyang Wang, Hoileong Lee, Xiaodi Pu, Yuanming Lai, Yiming Ma}
\date{}

\let\svthefootnote\thefootnote
\newcommand\blfootnotetext[1]{%
  \let\thefootnote\relax\footnote{#1}%
  \addtocounter{footnote}{-1}%
  \let\thefootnote\svthefootnote%
}

\let\svfootnotetext\footnotetext
\renewcommand\footnotetext[2][?]{%
  \if\relax#1\relax%
    \ifnum\value{footnote}=0\blfootnotetext{#2}\else\svfootnotetext{#2}\fi%
  \else%
    \if?#1\ifnum\value{footnote}=0\blfootnotetext{#2}\else\svfootnotetext{#2}\fi%
    \else\svfootnotetext[#1]{#2}\fi%
  \fi
}

\begin{document}
\maketitle

\begin{abstract}
\noindent In clinical medicine, rats are commonly used as experimental subjects. However, their estrous cycle significantly impacts their biological responses, leading to differences in experimental results. Therefore, accurately determining the estrous cycle is crucial for minimizing interference. Manually identifying the estrous cycle in rats presents several challenges, including high costs, long training periods, and subjectivity. To address these issues, this paper proposes a classification network — Spatial Long-distance EfficientNet (SLENet). This network is designed based on EfficientNet, specifically modifying the Mobile Inverted Bottleneck Convolution (MBConv) module by introducing a novel Spatial Efficient Channel Attention (SECA) mechanism to replace the original Squeeze Excitation (SE) module. Additionally, a Non-local attention mechanism is incorporated after the last convolutional layer to enhance the network's ability to capture long-range dependencies. The dataset used 2,655 microscopic images of rat vaginal epithelial cells, with 531 images in the test set. Experimental results indicate that SLENet achieved an accuracy of $96.31 \%$, outperforming baseline EfficientNet model (94.2\%). This finding provide practical value for optimizing experimental design in rat-based studies such as reproductive and pharmacological research, but this study is limited to microscopy image data, without considering other factors like temporal patterns, thus, incorporating multi-modal input is necessary for future application.
\end{abstract}

\noindent Index Terms-Rat estrous cycle classification, EfficientNet, Attention mechanism, Neural networks, Deep learning

\footnotetext{ 
\noindent The first two authors contributed equally to this work.\\
\indent Qinyang Wang, Yuanming Lai are with the School of Mechanical and Electrical Engineering, Chengdu University of Technology, Chengdu 610059, China (e-mail: \href{mailto:qyw333517@163.com}{qyw333517@163.com}, \href{mailto:laiyuanming19@cdut.edu.cn}{laiyuanming19@cdut.edu.cn})

Hoileong Lee is with the Faculty of Electronic Engineering \& Technology, University Malaysia Perlis, Arau, Malaysia (e-mail: \href{mailto:hoileong@unimap.edu.my}{hoileong@unimap.edu.my}).

Xiaodi Pu is with the Reproductive Section, Huaihua City Maternal and Child Health Care Hospital, Huaihua, China (email: \href{mailto:xiaodipu2022@163.com}{xiaodipu2022@163.com}).

Yiming Ma is with the CSG PGC Energy Storage Research Institute, China Southern Power Grid Co.Ltd, GuangZhou, China (email: \href{mailto:xmayiming@ieee.org}{xmayiming@ieee.org}).
}

\section*{I. Introduction}
Laboratory rodents, particularly rats, are widely used animal models in many fields of study related to human disease [1], [2], drug development [3], and genetic function [4]. In 2015, the National Institutes of Health (NIH) in policy NOT-OD-15-102 highlighted the significance of considering sex as an experimental variable to assess its impact on outcomes [5]. Similarly, the Canadian Institutes of Health Research (CIHR), in its Sex and Gender-Based Analysis Policy, emphasizes that sex and gender can influence disease susceptibility, response to pharmacological treatments, and patterns of healthcare utilization [6]. As a result, the number of medical experiments involving female rats has gradually increased [7],[8]. However, the estrous cycle of animals (including rats) significantly affects gene expression [9], protein levels [10], [11], [12], [13], behavior [14], [15], and drug responses [16], [17], [18], leading to substantial differences in experimental results. For example, Kaustubh et al. [19] employed rat as an animal model to investigate the impact of different stages of the estrous cycle on the oral bioavailability of Genistein, an active anticancer compound. Their findings showed that, with higher estrogen levels enhancing hepatic metabolism and consequently reducing systemic bioavailability. Another study by Lovick et al. [20] examined the influence of the estrous cycle on anxiety-related behaviors and pharmacological responses in female rats, with the aim of informing strategies for alleviating menstrual-related anxiety disorders in women. The study revealed that female rats exhibited heightened anxiety-like behaviors and fear responses during the diestrus phase. Moreover, the anxiolytic efficacy of benzodiazepines was more pronounced during the proestrus phase, whereas selective serotonin reuptake inhibitors (SSRIs) were more effective during diestrus. According to these studies, we can notice that misclassify the estrous cycle of rat can lead to erroneous conclusions, such as attributing hormone-induced behavioral variations to sex differences, or result in flawed assessments of drug metabolism and therapeutic efficacy, potentially translating into clinical risks-such as inappropriate drug selection, dosing inaccuracies, and adverse treatment outcomes.Therefore, accurately determining the estrous cycle of rats is crucial for conducting experiments.

The estrous cycle in rats mainly consists of four stages: Proestrus (P), Estrus (E), Metestrus (M), and Diestrus (D), with a typical cycle lasting 4-5 days [21]. Various methods are currently used to classify the stages of the estrous cycle in rats, with the most common being the identification of the types, shapes, numbers, sizes, and proportions of vaginal smear cells [22], [23], [24], [25]. The characteristics of each stage are as follows: the D stage is marked by the presence of a large number of leukocytes and small number of nucleated cells in the smear; the P stage contains nucleated epithelial cells and a small number of keratinized cells but no leukocytes; the E stage only contains anucleated keratinized epithelial cells, with no leukocyte cells; the M stage is marked by mixture of keratinized cells, nucleated epithelial cells, and leukocytes [26], [27]. Figure 1 shows microscopic images of vaginal smear cells at the four stages of the rats' estrous cycle.

In experiments, manually identifying the estrous cycle of rats is a commonly used and effective method. Hubscher et al. [27] used a method called PAP to stain vaginal smears of rats, quantifying the different cell populations throughout the cycle and providing guidelines for distinguishing the stages of the estrous cycle. However, several issues remain: (1) The efficiency of manual classification is low; (2) There is subjectivity in the classification results, leading to variations in outcomes; (3) The diverse types of cells at different stages make it challenging to distinguish them, and misidentification is likely if the examiner lacks specially trained.

\includegraphics[max width=200 pt, center]{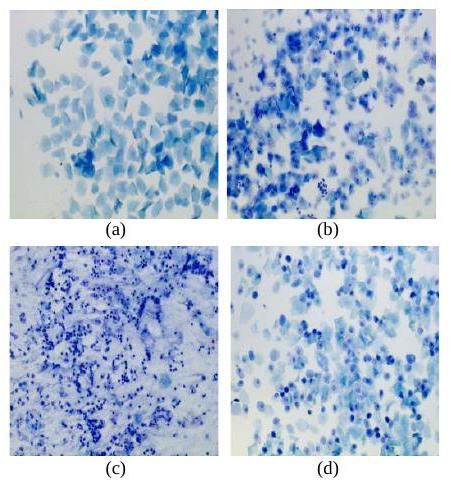}

\noindent Fig. 1. Stages of the estrous cycle include the (a) proestrus, (b) estrus, (c) metestrus, and (d) diestrus.\\

In recent years, deep learning [28], [29], particularly research methods based on convolutional neural networks (CNN) have been widely applied in the field of medical image processing [30], [31], [32], [33]. By adjusting the convolutional kernel size, CNN extracts local features at various scales [34], and its hierarchical structure captures finer-grained representations. More importantly, the design of parallel computing aligns perfectly with the operational logic of GPUs [35], [36]. Therefore, CNN-based models have achieved excellent performance in the field of medical image processing. For instance, HAQ M A et al. [37] developed a classification model called DCNNBT by modifying and optimizing the EfficientNet network for brain tumor images, using a large dataset of brain MRI images for training, ultimately achieving an accuracy of $99.18 \%$. In the same task, Liao et al. [38] proposed a model called GraphMriNet, achieved an average accuracy of $99.92 \%$ on four open datasets. Walid et al. [39] proposed a lightweight network model for classifying multi-modal medical images based on traditional CNN architecture; this model achieved classification accuracies of $92.7 \%, 91.1 \%, 100 \%$, and $100 \%$ in the task of classifying medical images from ultrasound, Xrays, CT, and MRI datasets under direct training.

Using CNN-based model to assess the estrous cycle in rats has achieved some promising results. For example, Kyohei et al. [40] utilized a VGG16-based network, SECREIT, to identify rat vaginal cells in microscopic images to classify their estrous cycle. They compared the performance of this model with experienced examiners and found that SECREIT achieved an accuracy of $93.3 \%$, surpassing the accuracy of the examiners. Wolcott et al. [41] proposed EstrousNet, another CNN-based model to deal the same task, they used ResNet50 architecture with transfer learning, achieving an average accuracy of $88.9 \%$. Babaev et al. [42] proposed ODES, a twostage estrous cycle classification framework that first employs YOLOv8 to detect and categorize individual vaginal epithelial cells, followed by a rule-based algorithm to classify the estrous stages based on cell type and proportions, achieving up to $88 \%$ accuracy. However, these models still present some limitations, for example, SECREIT and EstrousNet use two fully connected layers (with 500 and 3 nodes) may excessively compress the high-dimensional convolutional features, which can result in the loss of complex information such as cell type proportions and spatial relationships. ODES decouples feature extraction and stage classification (e.g. high-level stage inference is based solely on discrete cell counts), which may limit the model's ability to learn hierarchical or contextdependent features, and its performance deteriorates in cases of cell clustering, low contrast, or ambiguous transitional samples, where features are essential but underutilized.

Although existing studies have attempted to classify the estrous cycle stages of female rats using deep learning methods, most approaches still lack the capability to extract morphological and cytological features specific to vaginal smear images. Moreover, there is currently a lack of customized models and attention mechanisms that can effectively integrate channel features, spatial features, and global information of mouse vaginal cell images. As a result, existing methods often overlook some unique challenges, such as cell distribution patterns, ambiguous boundaries between categories, and staining artifacts. Therefore, to fill the gap, further enhance the performance of the deep learning model in the classification of the rat estrous cycle, we propose a novel multi-scale model that combines feature fusion and global attention — SLENet.

This model is designed to analyze microscopic images of rat vaginal epithelial cells and classify the four stages of the estrous cycle, providing classification results as a reference to assist researchers in their assessments.

The main contributions of this work are summarized as follows:
\begin{enumerate}
  \item We construct a dataset of 2655 stained images of rat vaginal epithelial cells, which integrates veterinary science, biomedical research, and computer vision, providing valuable data support for applying deep learning methods to estrous cycle classification tasks.
  \item We proposed SLENet, a multiscale medical image classification network that integrates a novel attention mechanism, SECA. SECA improves upon conventional attention approaches by jointly capturing channel and spatial features while maintaining computational efficiency, providing a solution for medical image classification in a specific domain.
  \item We further enhance the model's ability to capture global information by integrating a Non-local module after the final convolutional layer. This design enables the network to effectively model long-range dependencies, addressing the challenge of insufficient global feature modeling capability in specific medical image classification tasks.
\end{enumerate}

\section*{II. Materials \& Methods}
\section*{A. Dataset}
In this study, vaginal exfoliative cell staining was used to obtain microscopic images of the estrous cycle in rats. The specific procedure was as follows: A medical cotton swab moistened with saline was used to collect cells from the vaginal wall of the rat, which were then evenly smeared onto a glass slide. After air drying, the slides were stained with $0.2 \%$ methylene blue for approximately 15 minutes, followed by rinsing with water and air drying again before sealing for preservation. The experimental rats were injected with cyclophosphamide and leucovorin. Cyclophosphamide was used to induce immunosuppression, providing an experimental model to study the effects of immunosuppression on reproduction. Additionally, cyclophosphamide can cause alterations or disruptions in the estrous cycle of rat, which helps in collecting cell images representing various cycle stages with more diverse cellular morphology. Leucovorin was administered as a protective agent to mitigate the toxic side effects caused by cyclophosphamide, thereby ensuring the animals maintained a basic level of health during the experiment. Data were collected daily over a period of four weeks.

The annotator in this study is a researcher with a veterinary science and animal physiology background, with over six years of experience in laboratory animal management and reproductive physiology. Based on expert domain knowledge [27][43], the estrous cycle stages-estrus (E), proestrus (P), metestrus (M), and diestrus (D)-were annotated by analyzing the morphology, quantity, and proportion of cornified epithelial cells, leukocytes, and nucleated epithelial cells in the images. The research team conducted spot checks and reviews on a subset of the samples, no significant discrepancies were observed. Finally, the estrous cycle was classified into four phases (P, E, M, D), yielding 646, 672, 670, and 667 images, respectively. The dataset used in this study can be accessed from the corresponding author for reasonable request.

\noindent \includegraphics[max width=\textwidth, center]{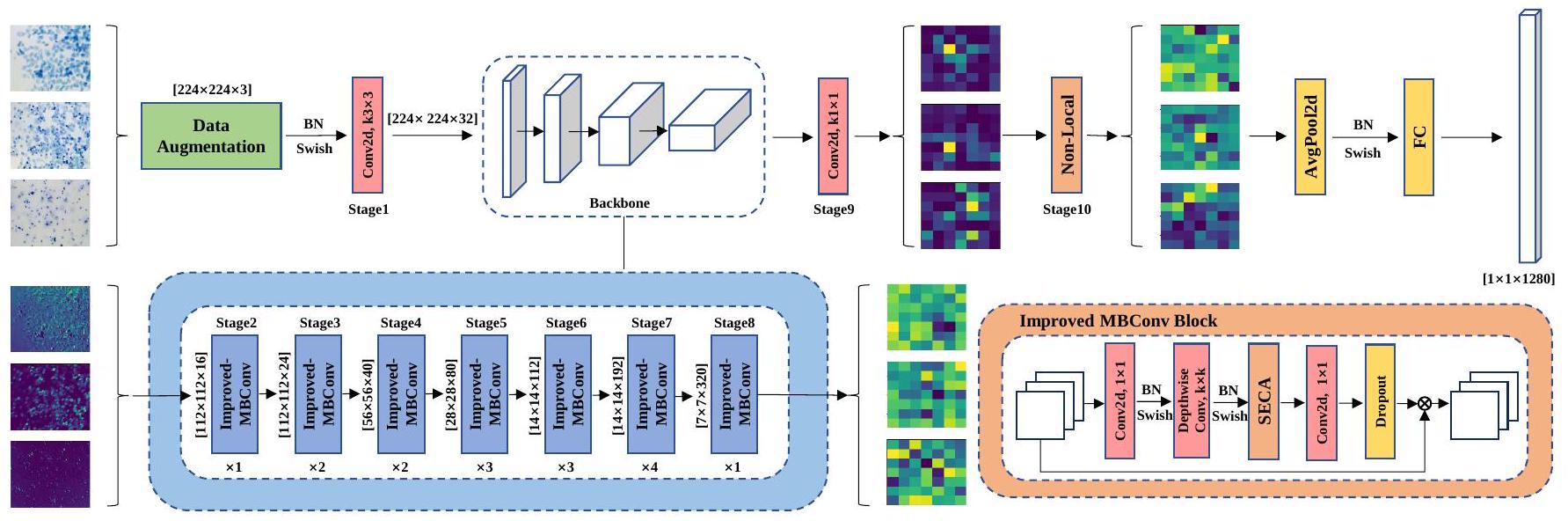}

\noindent Fig. 2. The overall architecture of SLENet

\section*{B. Data Preprocessing}
Data preprocessing aims to ensure a consistent format and enhance the comprehensiveness of the dataset, improving the accuracy and robustness of the model. In the original dataset, some images exhibited noticeable differences in brightness, which may have been caused by uneven smear thickness during slide preparation or cell overlap due to high cell density. Upon comparison, the brightness variation was found to be within $20 \%$. Moreover, due to differences in cellular structures, appropriate contrast adjustments were necessary to extract richer detail from the images. Therefore, to enhance the model's generalization capability, we compared the variations in the original images and conducted multiple rounds of experiments. Based on these results, we selected the best-performing adjustment strategy-randomly increasing or decreasing brightness by $15 \%$ and contrast by $10 \%$ as part of the image preprocessing process. Additionally, data augmentation methods like flipping and random rotation are applied to expand the dataset, allowing the results to have a certain degree of generalization. All images are converted into JPG format, which is suitable for processing by convolutional networks.

Before training, the images in the original dataset needed to be rescaled to match the input requirements ($224 \times 224$) of SLENet. In this process, interpolation methods were used to reasonably estimate and reconstruct the values of new pixels, thereby preserving the visual quality and information content of the images after scaling. Commonly used simple interpolation methods present various issues. For instance, nearest-neighbor interpolation often results in a strong aliasing effect, with jagged edges and significant detail loss in the cell images. Although bilinear interpolation offers some improvement, it still lacks sufficient smoothness for cellular images, potentially causing blurred cell boundaries. Other methods, such as Lanczos or area-based interpolation, may theoretically produce better results but are computationally intensive and more sensitive to noise, which can lead to edge artifacts after scaling and compromise image quality.

In comparison, bicubic interpolation offers a better balance by preserving image detail quality while maintaining relatively low computational complexity. Therefore, this study adopted bicubic interpolation as the method for image scaling. The interpolation function defined as follows:

\[
W(x)=\left\{\begin{array}{lc}
(a+2)|x|^{3}-(a+3)|x|^{2}+1 & \text { for }|x| \leq 1  \tag{1}\\
a|x|^{3}-5 a|x|^{2}+8 a|x|-4 a & \text { for } 1<|x|<2 \\
0 & \text { otherwise }
\end{array}\right.
\]

\noindent where x represents the absolute distance from the target pixel to its neighboring points, and a is a constant typically set to -0.5 or 1 . This function is used to calculate the weights of the target pixel relative to the 16 surrounding neighbor pixels, and the target pixel value can then be computed using the following formula:

\begin{equation*}
I(x, y)=\sum_{i=-1}^{2} \sum_{j=-1}^{2} W\left(x-x_{i}\right) \cdot W\left(y-y_{i}\right) \cdot I\left(x_{i}, y_{i}\right) \tag{2}
\end{equation*}

\noindent by using bicubic interpolation allows the compressed images retain more information from the original images, resulting in more accurate model outputs.

To train and evaluate the model's performance, the original dataset was divided into training, validation, and test sets at a ratio of 6:2:2 (1,593 images for training, 531 for validation, and 531 for testing). A random sampling strategy was employed during the splitting process to ensure independence and generalizability across the training and evaluation phases.

\section*{C. EfficientNet Model}
The EfficientNet model, proposed by Tan and Le in 2019 [44], is designed based on CNN. The network architecture is automatically optimized using Neural Architecture Search (NAS) algorithms, balancing computational efficiency and accuracy. EfficientNet introduces a compound scaling technique that utilizes a fixed compound factor $\Phi$ to allow simultaneous adjustments of the network's depth, width, and resolution. Compared to the traditional single-dimensional scaling techniques used in conventional network models, this approach more evenly increases model capacity and utilizes resources more effectively, enhancing model performance while maintaining computational efficiency, as described by formula 3:

\begin{gather*}
\text { Depth: } D=\alpha^{\Phi} \\
\text { Width: } W=\beta^{\Phi} \\
\text { Resolution }: R=\gamma^{\Phi}  \tag{3}\\
\text { s.t. } \alpha \cdot \beta^{2} \cdot \gamma^{2} \approx 2
\end{gather*}

The core of this model is the Mobile Inverted Bottleneck Convolution (MBConv) module, which incorporates the Squeeze-and-Excitation (SE) channel attention mechanism. The design of this module is similar to that of MobileNetV2 and utilizes an inverted residual structure, providing strong feature extraction capabilities. The EfficientNet model is composed of 16 stacked MBConv modules, along with 2 convolutional layers, 1 global average pooling layer, and 1 fully connected layer.

\section*{D. Improved MBConv Module}
Attention mechanisms are widely applied in computer vision[45], primarily functioning to mimic human attention by extracting key information and discarding irrelevant data, thus enhancing model performance. The original MBConv module incorporates the Squeeze-and-Excitation (SE) attention mechanism, which compresses spatial information through a Squeeze layer to generate channel descriptors. An Excitation layer then learns channel weight coefficients, and is applied to the original channel feature maps to improve network performance.

The classification of the estrous cycle does not rely on single dimension, it is more about a combination of features such as cell morphology, types and density. The channel attention mechanism SE can effectively differentiate the staining patterns of different cells, however, the average pooling function within the SE module compresses the spatial information of the image, which may diminish the attention to highdensity cell clusters or characteristic morphological regions in microscopic images, potentially overlooking some important local spatial information.

In this study, the SE attention module in the MBConv module is replaced by a novel attention mechanism called SECA. The specific structure is illustrated in Figure 3. This attention mechanism utilizes local 1D convolutions instead of fully connected layers, reducing computational complexity compared to the SE module. Additionally, the new module introduces two convolutional layers with a scaling ratio of 4 and a kernel size of 7. A Sigmoid activation function is employed to learn the spatial weights of the images, which are then applied to the output. This mechanism effectively scaling and extracting spatial information from the input feature maps, allowing MBConv not only capture the channel information, but also preserve spatial information, enhancing the model's focus on locally salient features and thereby improving its accuracy.

\section*{E. Non-Local Attention Mechanism}
The convolution operations used to process image data present local connectivity, meaning that the correlations of features extracted through convolution are limited to local regions. In microscopic images of rat cell smears, blank areas can occupy a significant portion of the image pixels. As a result, traditional convolutional layers with kernel sizes of $3 \times 3$ or $5 \times 5$ can effectively handle local information, but may lead to the majority of processing results being irrelevant to the task, since these blank regions do not contain cellular information.

\noindent \includegraphics[max width=\textwidth, center]{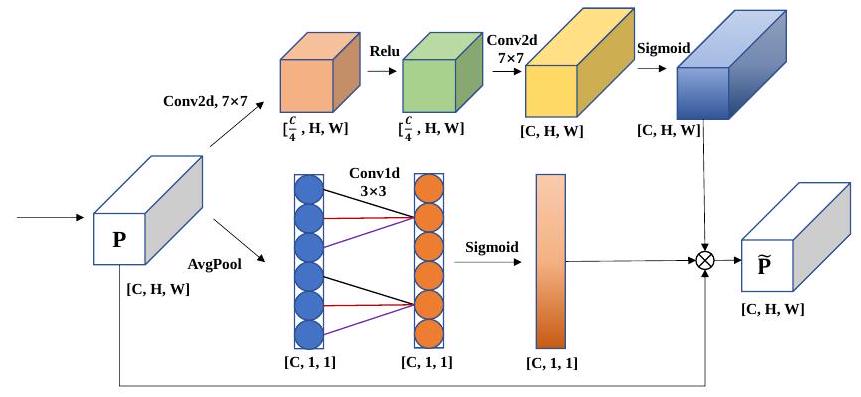}

\noindent Fig. 3. The structure of SECA \\

When classifying the estrous cycle in rats, it is essential not only to focus on the details of individual cells to determine their types and local spatial features but also to consider the overall quantity and distribution patterns of the cells, which means that global features in the images should be emphasized. Relying solely on traditional convolution operations may overlook these critical aspects, resulting in decreased model accuracy.

To address the issue of missing global information, this study introduces a Non-Local attention mechanism to capture longer-distance dependencies within the image. This mechanism utilizes self-attention to compute interactions between any two positions in the image, effectively extracting long-distance dependencies. The specific calculation is illustrated by formula 4:

\begin{align*}
& Q=X W_{q}, K=X W_{k}, V=X W_{v} \\
& F_{A}(Q, K, V)=\delta\left(\frac{Q K^{T}}{\sqrt{d_{k}}}\right) V \tag{4}
\end{align*}

\noindent where the $W_{q}, W_{k}, W_{v}$ represent the parameter matrices, while X is the input matrix. $F_{A}$ is the attention layer, and $\delta$ is the Softmax activation function; $d_{k}$ is the dimension of k . This mechanism effectively expands the convolutional kernel to match the size of the image, allowing for the capture of more global information and improving the classification accuracy of the model. In this study, the Non-Local attention mechanism is introduced after the final convolutional layer of SLENet, and the output is directly fed into the pooling layer, providing richer global information for subsequent classification. The specific structure is illustrated in Figure 4.\\
\noindent \includegraphics[max width=\textwidth, center]{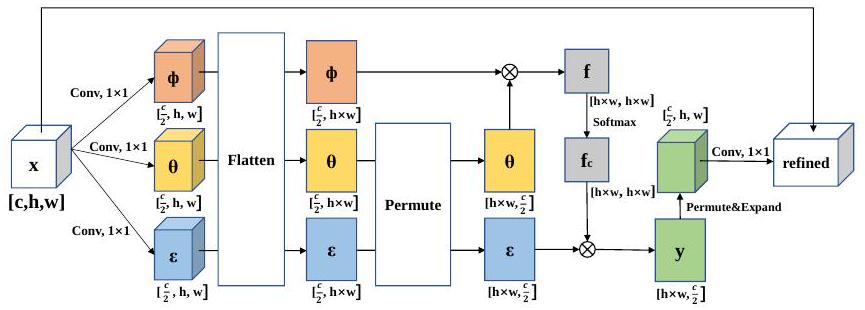}

\noindent Fig. 4. The structure of Non-Local

\section*{F. Experimental Environment}
To achieve optimal performance, this study determined reasonable ranges for key model hyperparameters-such as learning rate, batch size, and dropout rate before the fully connected layer based on prior related literature and experience in the field of medical image classification. A grid search strategy was then employed to evaluate various combinations of these hyperparameters. Each combination was trained under identical conditions, and the classification accuracy on the validation set was recorded. The best-performing set of hyperparameters was selected as the default configuration for this study: a learning rate of 0.01, batch size of 16. Additionally, we incorporate dropout layers as a regularization\\
technique to address the risk of overfitting due to irrelevant or noisy features. By using the same optimization method, we confirmed the optimal dropout rate was 0.2 . In addition, an early stopping mechanism was introduced during training, using validation accuracy as the evaluation criterion. The patience was set to 20 , and the optimal number of training epochs was determined to be 130.

Regarding optimizer selection, we compared commonly used optimizers in image classification tasks, including SGD, standard Adam, and AdamW. Validation accuracy was used for evaluation, and AdamW was found to perform the best in this study. This is because AdamW improves parameter regularization by decoupling weight decay from the momentum update, thus avoiding the issues present in traditional Adam where L2 regularization is entangled with adaptive updates-an improvement especially beneficial when training deeper networks. Moreover, the cellular images in the dataset exhibit heterogeneity (e.g., variations in morphology and density), and compared to SGD, AdamW can adaptively adjust the step size based on different gradient scales, leading to faster convergence and better capture of fine-grained features, making it more suitable for this task.

The experiments related to this study were conducted on a Windows 11 operating system, with a CPU model of $\operatorname{Intel}(\mathrm{R})$ Core(TM) i5-13600KF. A graphics processing unit (GPU) was utilized to accelerate the model training efficiency, and all experiments were carried out on a NVIDIA Corporation 4070 GPU for model training and testing. The specific model settings are detailed in Table I.

\begin{table}[ht]
\centering
\caption{The hyperparameters of Model training}
\label{tab:training-config}
\begin{tabular}{cc}
\hline
Parameters & Operator \\
\hline
Learning rate & 0.01 \\
Epoches & 130 \\
Batch size & 16 \\
Image size & $224 \times 224$ \\
Loss function & CrossEntropyLoss \\
Optimization algorithm & Adamw \\
\hline
\end{tabular}
\end{table}

\section*{III. Result \& DISCUSSION}
\section*{A. Performance Indicators}
When evaluating network models, specific assessment criteria are typically required to quantify and compare the model's performance. In this study, accuracy, precision, recall, and F1 score were selected as metrics to evaluate the model's performance. The confusion matrix serves as an analytical tool that visualizes the model's classification of the test samples, enabling the calculation of evaluation metrics based on these visualized data.

Accuracy is a key metric that measures how correctly a model performs its classification tasks. The specific formula for accuracy is given as follows:

\begin{equation*}
\text { Accuracy }=\frac{T P+T N}{T P+T N+F P+F N} \tag{5}
\end{equation*}

\noindent Precision reflects the proportion of truly positive samples among those that the model predicts as positive. The specific formula for precision is given as follows:

\begin{equation*}
\text { Precision }=\frac{T P}{T P+F P} \tag{6}
\end{equation*}

\noindent Recall reflects how many actual positive samples are predicted as positive. The specific formula for recall is given as follows:

\begin{equation*}
\text { Recall }=\frac{T P}{T P+F N} \tag{7}
\end{equation*}

\noindent Where TP represents true positives, reflecting the number of positive samples correctly identified by the model; TN represents true negatives, reflecting the number of negative samples correctly identified; FP represents false positives, reflecting the number of negative samples incorrectly identified as positive; FN represents false negatives, reflecting the number of positive samples incorrectly identified as negative.

The F1 Score balances the model's Precision and Recall, taking into account both the completeness (Recall) and the correctness (Precision) of the model's predictions. This provides a more comprehensive evaluation metric. The specific formula for the F1 Score is given as follows:

\begin{equation*}
F 1 \text { Score }=\frac{2 *(\text { Precision } * \text { Recall })}{\text { Precision }+ \text { Recall }} \tag{8}
\end{equation*}

\section*{B. Experimental Results}
For the established training and validation sets, comparative experiments were conducted using seven models: EfficientNet, ResNet18, ResNet34, ResNet50, VGG16, MobileNetV2, GoogleNet, DenseNet and the proposed SLENet. Additionally, we introduced the Vision Transformer (ViT) as a comparative model to explore the applicability of the transformer architecture in this task. Although transformer-based models have achieved excellent performance in various computer vision tasks, in our study, ViT showed a higher training loss and significantly lower validation accuracy compared to CNN-based models, as shown in Figure 5 and 6. This may because transformer architecture generally lacks the inductive biases inherent to CNNs, such as local receptive field and translation invariance, which may limit its ability to extract effective features from relatively small datasets. In our study, due to the limited number of experimental animal samples, the dataset was insufficient to support the effective training of transformer-based models. Furthermore, the self-attention mechanism employed in transformers has a quadratic computational complexity $\left(O\left(n^{2}\right)\right)$, leading to higher computational requirements during training. In contrast, convolutional operations are more efficient, making CNNs more suitable for conditions with limited computational resources. Therefore, the following analysis will focus on the performance of CNNbased models for this task.

Figure 5 indicates that the fluctuations of EfficientNet and ResNet50 in the early training stages are more pronounced compared to the other models. After approximately 100 epochs, the validation accuracy of all models shows reduced fluctuation and begins to converge. Notably, SLENet demonstrates a clear advantage after about 60 epochs, achieving a validation accuracy exceeding $96 \%$. This indicates an improvement in the classification performance of SLENet compared to commonly used convolutional neural networks. Importantly, SLENet exhibits smaller fluctuations in the accuracy curve throughout the training process, suggesting better generalization ability and stability than the other models.

Figure 6 presents the loss curves for each model on the training set. It can be observed that EfficientNet has the highest loss in the first 50 training epochs, while ResNet18 and ResNet34 perform well with lower loss values compared to the other models. After about 120 epochs, the loss values for all models change minimally, indicating convergence. During the 120-130 epoch range, SLENet displays lower loss values compared to the other models.

To provide a more comprehensive evaluation of the effectiveness of SLENet compared to other convolutional neural networks in the classification task of the rat estrous cycle, we present the prediction results of each model on the test set using a confusion matrix, as shown in Figure 7.\\

\noindent \includegraphics[max width=\textwidth, center]{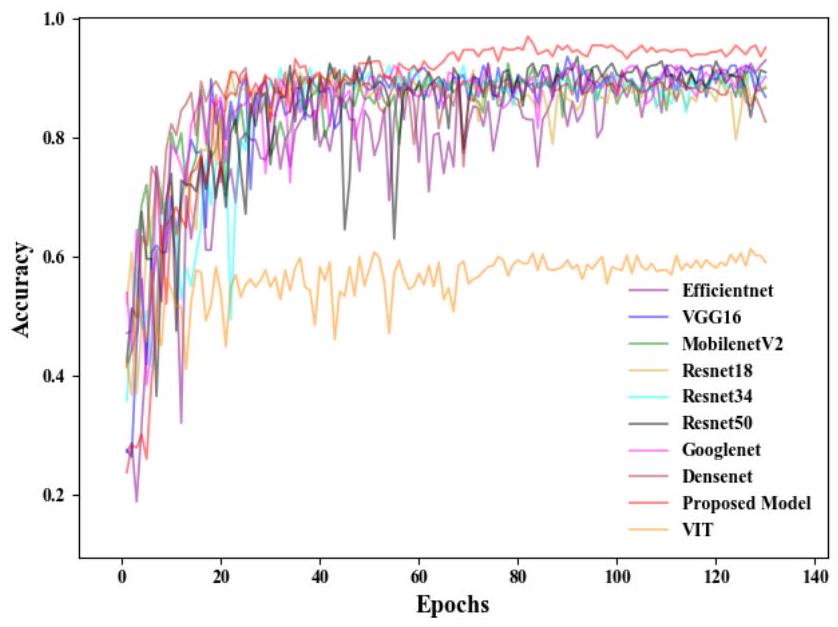}

\noindent Fig. 5. Validation accuracy of each model\\

The confusion matrix indicates that SLENet demonstrates the highest accuracy in identifying the estrous stages, correctly classifying all images from this phase. It also shows strong performance in recognizing the proestrus phase, with only 1 image misclassified as the diestrus phase. Additionally, there were 6 and 12 misclassifications for the diestrus and metestrus phases, respectively. Compared to other classification models, which had total misclassifications exceeding 20, SLENet shows its advantage.

Figure 9 shows normalized confusion matrix data in the form of a bar chart, as can be seen, the proposed model achieves highest accuracy across E, D, P stages, demonstrating a comparatively superior generalization and robustness. Notably, SLENet does not achieve the top accuracy in the M stage, it still maintains competitive performance, closely following the best-performing model, this slight drop may be attributed to the subtle and transitional nature of Metestrus, which poses greater challenges for all models.

To ensure the statistical reliability of our experimental results, we conducted each experiment five times with different random seeds under same environment, reporting the mean and 95\% confidence intervals for evaluation metric, and performed statistical significance testing between SLENet and other models, as shown in Table II-IV. Considering that precision and recall are often correlated in value, and F1 score provides a balanced measure between them, therefore, we applied paired t-tests to calculate the P-values of each class only based on the F1 score and report the average value as an overall indicator.\\

\noindent \includegraphics[max width=\textwidth, center]{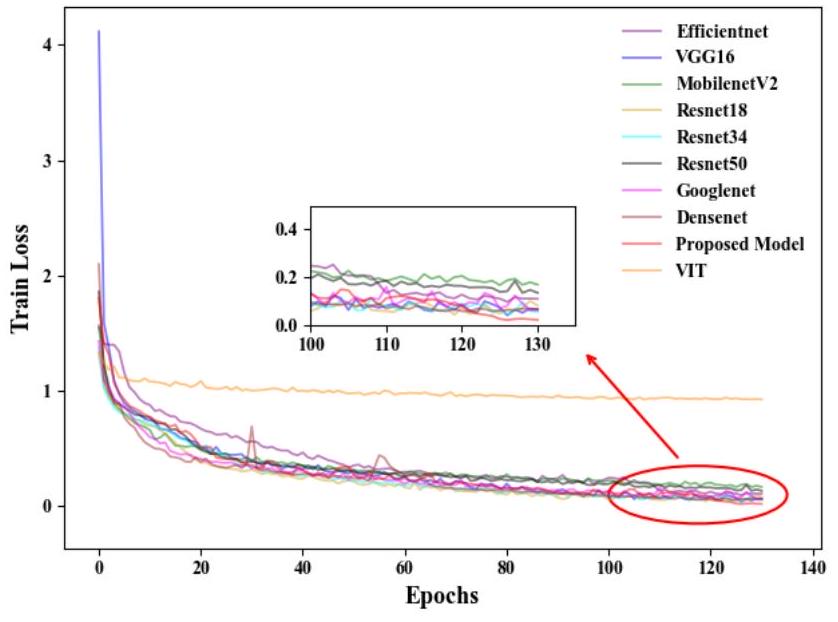}

\noindent Fig. 6. Training loss of each model\\

Based on the results, As shown in Figure 8, the average accuracy of SLENet is $96.31 \%$, which is the highest among these models. Table II-IV show that, SLENet's overall average value on Precision, Recall, and F1-score achieved $96.27 \%$, $96.30 \%$, and $96.26 \%$ respectively, with smaller confidence intervals ($3.65 \%, 5.76 \%$ and 4.18\%), indicating excellent classification accuracy and robustness. Notably, in some cases, although SLENet shows slightly lower average Precision and Recall compared to the best-performing model, it consistently exhibits the smallest confidence intervals. This indicates that its predictions are more stable and reliable. More importantly, SLENet achieves the highest F1-score in all four classes, which means it can balance Precision and Recall more effectively, demonstrating better overall performance in this task. Additionally, the results show that SLENet achieves statistical significance ($\mathrm{P}_{¡} 0.05$) when compared with most of these models. Although the comparison with EfficientNet results in a P-value of 0.13, which does not meet the significance threshold, the proposed model still outperformed EfficientNet numerically in all classes, showing an overall advantageous performance trend.

To further evaluate the performance of SLENet in this multiclass classification task, the Receiver Operating Characteristic (ROC) and Precision-Recall (PR) curves were generated. The ROC curves were constructed using one-vs-rest (OvR) strategy. As shown in Figure 10 and 11, the Area Under the Curve (AUC) and Average Precision (AP) for all classes in the figure exceed 0.99, showing that despite the overall accuracy implies a few errors in prediction, the model has excellent ranking capability and robust probability outputs, the few misclassifications did not significantly impact the model's ability to distinguish between classes or to correctly identify positive samples.\\

\noindent \includegraphics[max width=\textwidth, center]{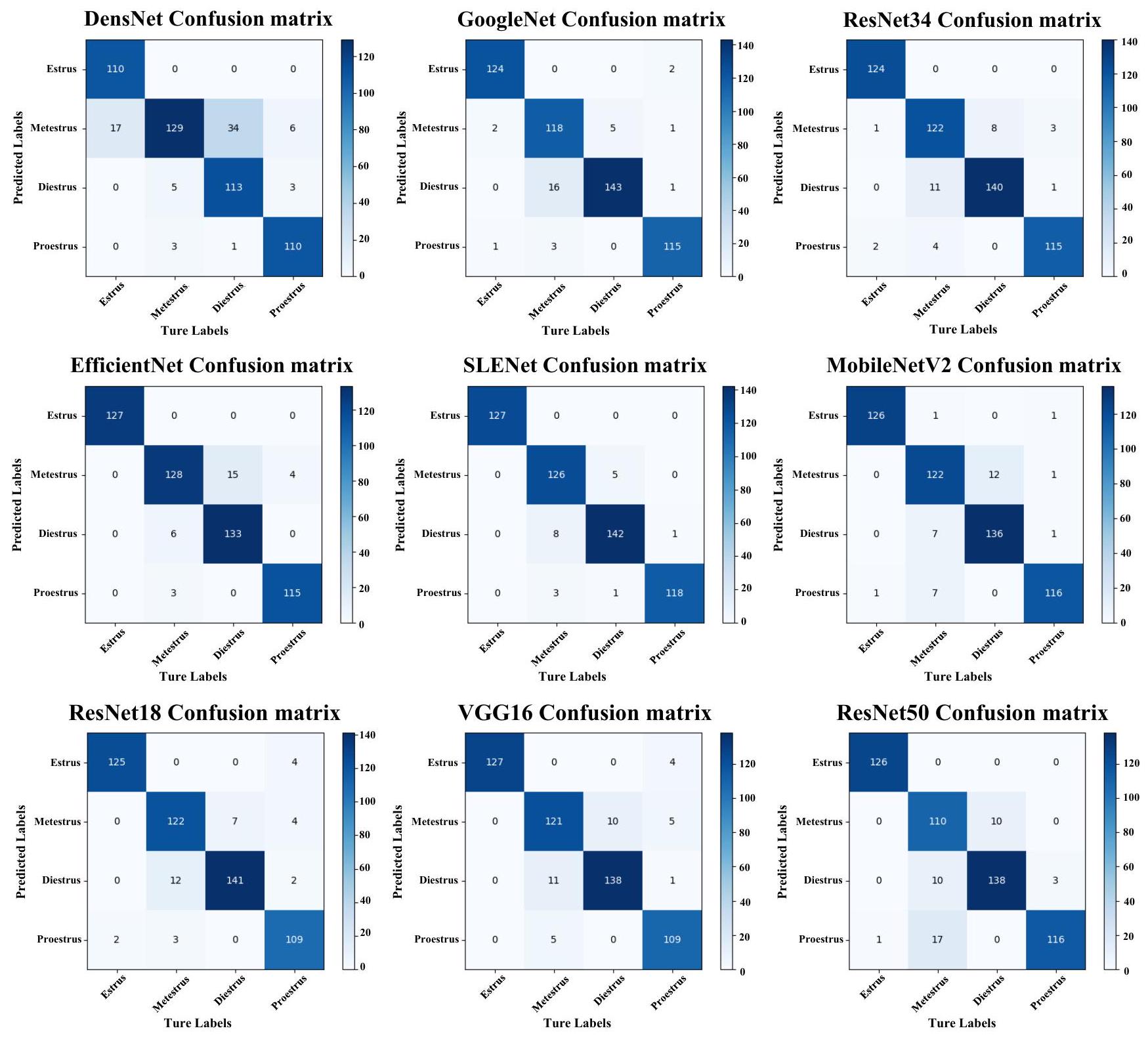}

\noindent Fig. 7. The confusion matrix of each model\\

Analyzing the result from a biological perspective, the estrous cycle in rat is a dynamic process, therefore, during the construction of the dataset, transitional phases are inevitable. During these periods, vaginal cytology often contains a diverse range of cell types in large quantities, resulting in images rich in detailed textures. Consequently, some issues such as cell overlap, blurred edges, and uneven staining may occur in the collected images. For these atypical images, in manual classification, experts may incorporate multidimensional information to make flexible judgments, such images can be classified as "transition phases" or "suspected stage", and multiple experts may review the samples to improve accuracy when it is necessary. However, network models are trained based on fixed labels, these factors present challenges for relatively simple models (e.g., those without integrated attention mechanisms), which may have limitations in effectively extracting such complex features, ultimately leading to performance differences.\\

\noindent \includegraphics[max width=\textwidth, center]{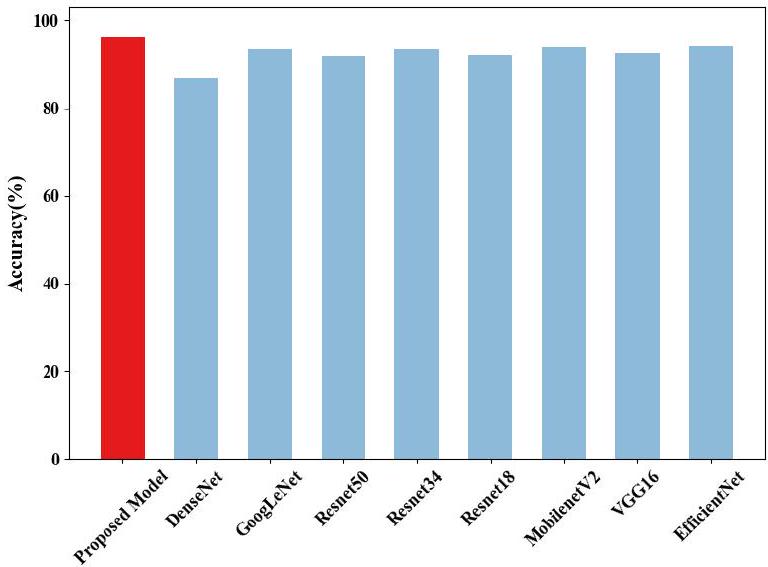}

 \noindent Fig. 8. The overall accuracy of each model\\

\noindent \includegraphics[max width=\textwidth, center]{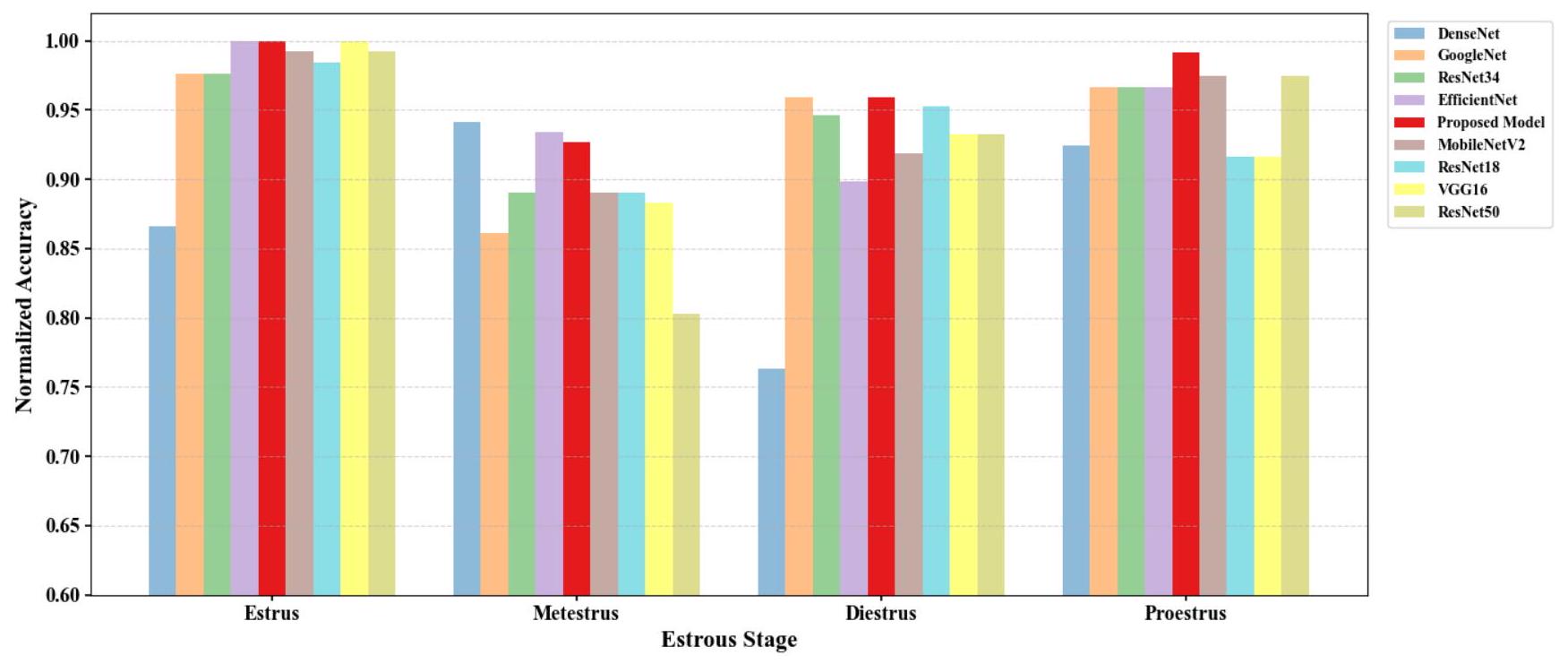}

\noindent Fig. 9. The bar chart of confusion matrix \\

According to the results, it is evident that the highest misclassification rates occur between the Metestrus (M) and Diestrus (D) stages. This is because the cytological composition during stages M and D is quite similar, with both contain a large number of leukocytes. For the model, subtle differences in leukocyte proportions are difficult to distinguish accurately. Additionally, we observed that the Proestrus (P) stage is often misclassified as Estrus (E). This is due to the gradual keratinization of nucleated epithelial cells on the vaginal smears during the late P stage, making the image features increasingly resemble those of the E stage and thus confusing the model. Lastly, it is noteworthy that stages E and D are generally well distinguished. This is because stage E is characterized by densely packed and orderly arranged keratinized cells, whereas stage D is dominated by small, round leukocytes. The distinct morphological features between these two stages make them relatively easier for the model to differentiate.

Generally, the experimental results suggest that SLENet is more suitable for the classification of specific medical images.

\begin{table}[ht]
\centering
\caption{Precision of Each Model}
\label{tab:model-precision}
\begin{tabular}{|l|l|l|l|l|l|}
\hline
Model & Estrus & Metestrus & Diestrus & Proestrus & Average \\
\hline
EffcientNet & $99.56 \pm 0.50$ & $87.91 \pm 3.83$ & $\mathbf{94.87} \pm \mathbf{0.84}$ & $95.36 \pm 2.13$ & $94.43 \pm 7.68$ \\
\hline
ResNet18 & $96.01 \pm 0.94$ & $89.49 \pm 4.68$ & $87.55 \pm 2.84$ & $95.28 \pm 2.00$ & $92.08 \pm 6.68$ \\
\hline
ResNet34 & $99.40 \pm 0.79$ & $91.36 \pm 1.28$ & $90.56 \pm 1.89$ & $93.34 \pm 1.76$ & $93.67 \pm 6.36$ \\
\hline
ResNet50 & $99.26 \pm 0.64$ & $90.14 \pm 2.64$ & $91.62 \pm 1.65$ & $86.51 \pm 3.69$ & $91.99 \pm 8.85$ \\
\hline
VGG16 & $96.63 \pm 0.66$ & $88.92 \pm 1.46$ & $89.65 \pm 3.43$ & $94.52 \pm 1.72$ & $92.43 \pm 5.96$ \\
\hline
MobileNetV2 & $98.21 \pm 0.96$ & $90.96 \pm 2.66$ & $93.83 \pm 2.58$ & $93.71 \pm 1.71$ & $93.93 \pm 4.92$ \\
\hline
GoogleNet & $98.48 \pm 0.69$ & $93.75 \pm 2.53$ & $87.46 \pm 2.84$ & $\mathbf{95.90} \pm \mathbf{1.99}$ & $93.90 \pm 7.49$ \\
\hline
DensNet & $98.38 \pm 1.57$ & $69.42 \pm 3.35$ & $93.50 \pm 2.54$ & $95.03 \pm 2.74$ & $89.08 \pm 21.11$ \\
\hline
SLENet & $\mathbf{99.69} \pm \mathbf{0.52}$ & $\mathbf{95.50} \pm \mathbf{1.21}$ & $\underline{94.85 \pm 0.57}$ & $95.05 \pm 1.16$ & $\mathbf{96.27} \pm \mathbf{3.65}$ \\
\hline
\end{tabular}
\end{table}

\begin{table}[ht]
\centering
\caption{Recall Of Each Model}
\label{tab:model-f1score}
\begin{tabular}{|l|l|l|l|l|l|}
\hline
Model & Estrus & Metestrus & Diestrus & Proestrus & Average \\
\hline
EffcientNet & $\underline{99.54} \pm 0.52$ & $91.59 \pm 2.22$ & $90.61 \pm 3.19$ & $95.62 \pm 1.80$ & $94.34 \pm 6.51$ \\
\hline
ResNet18 & $98.95 \pm 1.04$ & $85.84 \pm 5.30$ & $95.09 \pm 2.90$ & $89.48 \pm 3.65$ & $95.52 \pm 9.45$ \\
\hline
ResNet34 & $97.88 \pm 0.71$ & $86.27 \pm 3.15$ & $94.35 \pm 1.51$ & $95.90 \pm 2.13$ & $93.60 \pm 8.11$ \\
\hline
ResNet50 & $98.05 \pm 1.63$ & $79.84 \pm 4.65$ & $93.29 \pm 3.56$ & $\underline{98.10 \pm 0.91}$ & $93.32 \pm 13.72$ \\
\hline
VGG16 & $99.12 \pm 1.12$ & $84.83 \pm 5.31$ & $93.31 \pm 2.47$ & $91.20 \pm 0.64$ & $92.34 \pm 9.95$ \\
\hline
MobileNetV2 & $99.08 \pm 0.47$ & $88.65 \pm 1.61$ & $92.65 \pm 3.17$ & $95.61 \pm 2.66$ & $94.00 \pm 7.05$ \\
\hline
GoogleNet & $98.78 \pm 1.11$ & $83.42 \pm 4.41$ & $\mathbf{96.98} \pm \mathbf{1.30}$ & $94.84 \pm 1.29$ & $93.51 \pm 11.00$ \\
\hline
DensNet & $90.30 \pm 3.93$ & $\mathbf{91.92} \pm \mathbf{2.61}$ & $75.36 \pm 2.25$ & $89.98 \pm 2.51$ & $86.89 \pm 12.31$ \\
\hline
SLENet & $\mathbf{99.56} \pm \mathbf{0.81}$ & $\underline{91.65 \pm 1.39}$ & $\underline{95.25 \pm 1.15}$ & $\mathbf{98.75} \pm \mathbf{0.50}$ & $\mathbf{96.30} \pm \mathbf{5.76}$ \\
\hline
\end{tabular}
\end{table}

\begin{table}[ht]
\centering
\caption{F1 Score and P-value of Each Model(\% for F1 score $\uparrow$), P-values are shown in decimal; values smaller than 0.01 are presented in scientific notation}
\label{tab:model-accuracy}
\begin{tabular}{|l|l|l|l|l|l|l|}
\hline
Model & Estrus & Metestrus & Diestrus & Proestrus & Average & P-value ($\downarrow$) \\
\hline
EfficientNet & $99.55 \pm 0.38$ & $89.68 \pm 1.85$ & $92.67 \pm 1.34$ & $95.49 \pm 1.77$ & $94.35 \pm 6.69$ & 0.13 \\
\hline
ResNet18 & $97.46 \pm 0.39$ & $86.56 \pm 4.20$ & $91.48 \pm 2.41$ & $91.96 \pm 1.59$ & $91.87 \pm 7.09$ & $1.7 \times 10^{-3}$ \\
\hline
ResNet34 & $98.63 \pm 0.48$ & $88.69 \pm 1.22$ & $92.41 \pm 1.50$ & $94.59 \pm 1.08$ & $93.58 \pm 6.61$ & 0.038 \\
\hline
ResNet50 & $98.86 \pm 1.08$ & $86.63 \pm 6.07$ & $91.64 \pm 1.63$ & $92.53 \pm 3.72$ & $92.41 \pm 7.99$ & 0.043 \\
\hline
VGG16 & $98.29 \pm 0.34$ & $86.75 \pm 3.58$ & $91.42 \pm 1.95$ & $92.82 \pm 0.84$ & $92.32 \pm 7.56$ & 0.043 \\
\hline
MobileNetV2 & $98.64 \pm 0.50$ & $\underline{89.79} \pm 1.94$ & $93.22 \pm 2.40$ & $94.13 \pm 1.43$ & $93.95 \pm 5.80$ & 0.023 \\
\hline
GoogleNet & $98.47 \pm 0.46$ & $87.99 \pm 2.96$ & $91.96 \pm 1.86$ & $95.36 \pm 1.25$ & $93.45 \pm 7.17$ & 0.016 \\
\hline
DensNet & $94.14 \pm 2.05$ & $79.08 \pm 2.58$ & $83.20 \pm 1.59$ & $92.43 \pm 2.56$ & $87.21 \pm 11.53$ & $7 \times 10^{-4}$ \\
\hline
SLENet & $\mathbf{99.63} \pm \mathbf{0.35}$ & $\mathbf{93.53} \pm \mathbf{0.88}$ & $\mathbf{95.03} \pm \mathbf{0.78}$ & $\mathbf{96.86} \pm \mathbf{1.15}$ & $\mathbf{96.26} \pm \mathbf{4.18}$ & \textbf{--} \\
\hline
\end{tabular}
\end{table}

\noindent \includegraphics[max width=\textwidth, center]{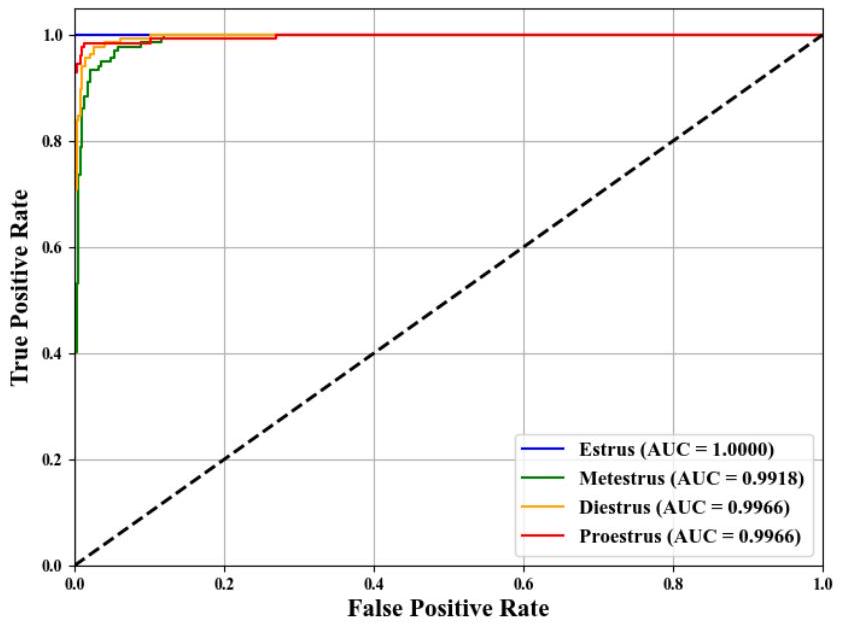}

\noindent Fig. 10. The ROC curve of SLENet\\

\noindent \includegraphics[max width=\textwidth, center]{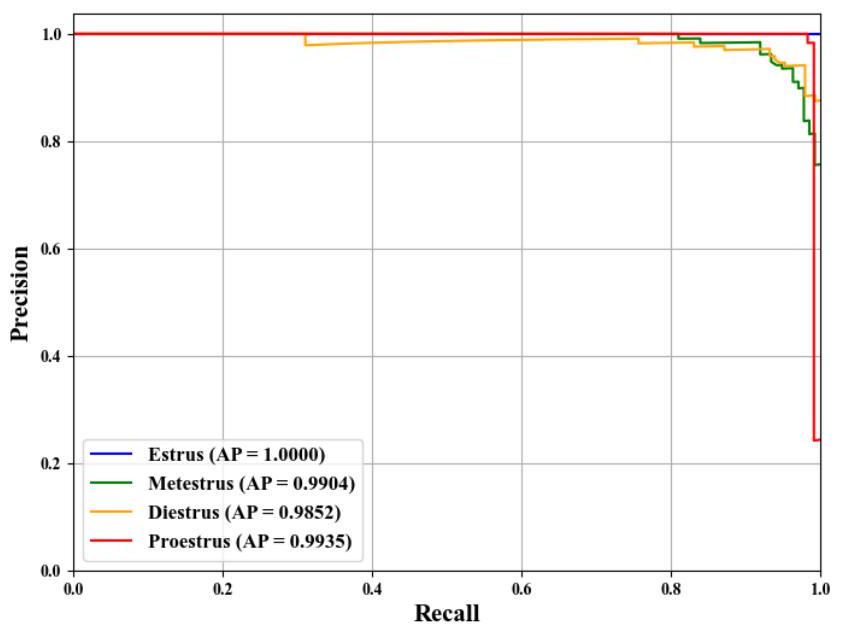}

\noindent Fig. 11. The PR curve of SLENet

\section*{C. Ablation Study}
This section uses ablation experiments to demonstrate the effectiveness of the modules introduced in the SLENet network. The control group consists of the SLENet network and EfficientNet. All other parameters and conditions are kept the same.

To ensure the reliability of the results, we use the same strategy to repeat the experiment five times and report the F1 score and overall accuracy as mean $\pm 95 \%$ confidence interval. Additionally, to evaluate whether SECA provides superior enhancement, we incorporated attention modules such as Convolutional Block Attention Module (CBAM) and Coordinate Attention (CA), which also emphasizes joint modeling of spatial and channel features, and calculated their performance. The specific results are shown in Table V.

\begin{table}[ht]
\centering
\caption{Classification performance table in ablation experiments (\%~$\uparrow$)}
\label{tab:ablation-classification}
\begin{tabular}{|l|l|l|l|l|l|l|}
\hline
Baseline & SECA & Non-Local & CBAM & CA & F1 & Accuracy \\
\hline
$\checkmark$ &  &  &  &  & $94.35 \pm 6.69$ & $94.20 \pm 0.99$ \\
\hline
$\checkmark$ & $\checkmark$ &  &  &  & $94.78 \pm 5.52$ & $\underline{94.55} \pm 0.86$ \\
\hline
$\checkmark$ &  & $\checkmark$ &  &  & $93.82 \pm 4.95$ & $94.12 \pm 0.56$ \\
\hline
$\checkmark$ &  & $\checkmark$ & $\checkmark$ &  & $91.29 \pm 9.36$ & $91.29 \pm 2.15$ \\
\hline
$\checkmark$ &  & $\checkmark$ &  & $\checkmark$ & $94.50 \pm 5.51$ & $94.39 \pm 0.65$ \\
\hline
$\checkmark$ & $\checkmark$ & $\checkmark$ &  &  & $\mathbf{96.26} \pm \mathbf{4.18}$ & $\mathbf{96.31} \pm \mathbf{0.43}$ \\
\hline
\end{tabular}
\end{table}

As we can see from the result, incorporating SECA alone can increase the mean F1-score and Accuracy, and reduce the confidence interval, but the overall improvement is quite limited. In contrast, when the Non-local module is introduced alone, both F1-score and Accuracy decrease. However, when SECA and Non-local are combined, the model achieves the best performance on both metrics (F1-score $=96.26 \%$, Accuracy $=96.31 \%$ ), with a reduction in confidence intervals, which means it not only enhances the model's predictive performance, but also improves its stability. This result can be explained as follows: In this task, vaginal smear microscopic images present both prominent local features (such as clusters of keratinized cell or leukocytes) and global features (such as the proportion and spatial arrangement of different cell types), relying solely on local details or global distribution can easily lead to miscalssification. For example, a local region might already show keratinized cells, but the overall distribution still resembles the previous stage; or certain areas is dominated by leukocyte, but the global proportion is not fully changed. SECA module enhances the model's capability to capture critical local features, improving its sensitivity to the details of the images. Additionally, Non-local module strengthens the model's ability to capture long-range dependencies, which is crucial for recognizing specific distribution patterns. Therefore, compared to introducing a single module, integrating both of them can improve the model's classification performance more effectively.

The results also show that substituting SECA module with the CA module leads to only minimal performance improvement, while substituting it with CBAM even degrades the model's performance. This indicates that the fusion methods of these modules are not suitable for discriminating features in this task, therefore, incorporating SECA is better to classify the estrous cycle of rats.

\section*{D. Complexity Analysis}
To assess the computational efficiency, we compared the inference time, number of parameters, and floating-point operations (FLOPS) between the baseline (EfficientNet) and SLENet. The inference time was measured over five runs and reported the average result under same environment to ensure consistency.

\begin{table}[ht]
\centering
\caption{Model complexity metrics}
\label{tab:model-complexity}
\begin{tabular}{c|ccc}
\hline
Model & Parameters & FLOPs & Inference time (ms) \\
\hline
Baseline & 4.01 M & 6.58 G & 32.32 \\
SLENet   & 14.19 M & 9.35 G & 34.58 \\
\hline
\end{tabular}
\end{table}

As shown in Table VI, compared to the baseline, SLENet shows an increase in model complexity, the number of parameters increases from 4.01 M to 14.19 M , which is about 3.5 times larger. Additionally, the FLOPs increases form 6.58 G to 9.35 G, with an increase of $42 \%$, both of them reflecting a higher computational complexity. This is primarily due to the introduction of SECA and Non-Local, which enhance the model's feature extraction capability but inevitably add extra parameters and computational cost.

However, despite the increase in both parameters and FLOPs, the inference time only rises from 32.32 ms to 34.58 ms , representing only $7 \%$ increase, indicating that the added computations are processed efficiently. Based on the previously discussed result, these increase in complexity is deemed acceptable and justified, reflecting a balanced design between accuracy and computational efficiency.

\section*{IV. Conclusion}
This paper presents SLENet, a medical image classification algorithm based on a CNN architecture, applied to the identification of the rat estrous cycle. The backbone of this algorithm is constructed from stacked improved MBConv modules, with the core component being the SECA attention mechanism, which integrates channel and spatial information from images. Additionally, the network employs a global attention mechanism before the fully connected layer to capture long distance dependencies in images, enhancing the model's receptive field and information interaction capabilities. Comparative analysis with current mainstream CNN architectures demonstrates that SLENet achieves state-of-the-art performance, with an average accuracy of $96.31 \%$.

However, despite SLENet's capability to accurately differentiate most categories, its effectiveness in distinguishing between the diestrus and metestrus phases remains limited due to a lack of capability to integrate multiple sources of information for classification. In future, incorporating temporal sequence information could help mitigate the interference introduced by transitional phases between estrous stages, enhancing the robustness of the model. Furthermore, we can consider extending the model to support multi-modal inputs, such as combining cytological images with hormone level, may further improve classification accuracy.

\section*{DATA AVAILABILITY STATEMENT}
Missing values, inconsistent data, and erroneous records may exist in the data set. These issues may affect the accuracy of analysis results. Requests to access these datasets should be directed to: \href{mailto:xiaodipu2022@163.com}{xiaodipu2022@163.com}.

\section*{Ethics statement}
The animal study was approved by Ethics Committee of Huaihua Maternal and Child Health Hospital. The study was conducted in accordance with the local legislation and institutional requirements.

\section*{REFERENCES}
[1] T. V. de Jong, Y. Pan, P. Rastas, D. Munro, M. Tutaj, H. Akil, C. Benner, D. Chen, A. S. Chitre, W. Chow, et al., 
``A revamped rat reference genome improves the discovery of genetic diversity in laboratory rats,'' 
Cell Genomics, vol. 4, no. 4, 2024.\\[0pt]
[2] M. Sato, S. Nakamura, E. Inada, and S. Takabayashi, 
``Recent advances in the production of genome-edited rats,'' 
International Journal of Molecular Sciences, vol. 23, 2022. [Online]. Available: 
\href{https://api.semanticscholar.org/CorpusID:247132843}{https://api.semanticscholar.org/CorpusID:247132843}\\[0pt]
[3] C. J. Morris, M. G. Rolf, L. Starnes, I. C. Villar, A. Pointon, H. Kimko, and G. Y. Di Veroli, 
``Modelling hemodynamics regulation in rats and dogs to facilitate drugs safety risk assessment,'' 
Frontiers in Pharmacology, vol. 15, 2024. [Online]. Available: 
\href{https://www.frontiersin.org/journals/pharmacology/articles/10.3389/fphar.2024.1402462}{https://www.frontiersin.org/journals/pharmacology/articles/10.3389/fphar.2024.1402462}\\[0pt]
[4] P. Soufizadeh, V. Mansouri, and N. Ahmadbeigi, 
``A review of animal models utilized in preclinical studies of approved gene therapy products: trends and insights,'' 
Laboratory Animal Research, vol. 40, 2024. [Online]. Available: 
\href{https://api.semanticscholar.org/CorpusID:269292786}{https://api.semanticscholar.org/CorpusID:269292786}\\[0pt]
[5] N. I. of Health, 
``NIH policy on sex as a biological variable,'' Website, 2015. [Online]. Available: 
\href{https://grants.nih.gov/grants/guide/notice-files/NOT-OD-15-102}{https://grants.nih.gov/grants/guide/notice-files/NOT-OD-15-102}\\[0pt]
[6] W. El-Shafai, A. A. Mahmoud, A. M. Ali, E.-S. M. El-Rabaie, T. E. Taha, A. S. El-Fishawy, O. Zahran, and F. E. A. El-Samie, 
``Efficient classification of different medical image multimodalities based on simple CNN architecture and augmentation algorithms,'' 
Journal of Optics, vol. 53, no. 2, pp. 775--787, 2024.\\[0pt]
[7] A. Freeman, P. Stanko, L. Berkowitz, N. Parnell, A. Zuppe, T. Bale, T. Ziolek, and C. Epperson, 
``Inclusion of sex and gender in biomedical research: Survey of clinical research proposed at the University of Pennsylvania,'' 
Biology of Sex Differences, vol. 8, Jun. 2017.\\[0pt]
[8] B. Prendergast and J. Liang, 
``Female rats are not more variable than male rats: A meta-analysis of neuroscience studies,'' 
Biology of Sex Differences, vol. 7, Dec. 2016.\\[0pt]
[9] S. Ray, R.-Y. Tzeng, L. M. DiCarlo, J. L. Bundy, C. Vied, G. Tyson, R. Nowakowski, and M. N. Arbeitman, 
``An examination of dynamic gene expression changes in the mouse brain during pregnancy and the postpartum period,'' 
G3 Genes—Genomes—Genetics, vol. 6, no. 1, pp. 221--233, Jan. 2016. [Online]. Available: 
\href{https://doi.org/10.1534/g3.115.020982}{https://doi.org/10.1534/g3.115.020982}\\[0pt]
[10] M. L. Zenclussen, P. A. Casalis, F. Jensen, K. Woidacki, and A. C. Zenclussen, 
``Hormonal fluctuations during the estrous cycle modulate heme oxygenase-1 expression in the uterus,'' 
Frontiers in Endocrinology, vol. 5, 2014. [Online]. Available: 
\href{https://www.frontiersin.org/journals/endocrinology/articles/10.3389/fendo.2014.00032}{https://www.frontiersin.org/journals/endocrinology/articles/10.3389/fendo.2014.00032}\\[0pt]
[11] J. Spencer, E. Waters, K. Bath, M. Chao, B. McEwen, and T. Milner, 
``Distribution of phosphorylated TrkB receptor in the mouse hippocampal formation depends on sex and estrous cycle stage,'' 
The Journal of Neuroscience: The Official Journal of the Society for Neuroscience, vol. 31, pp. 6780--6790, May 2011.\\[0pt]
[12] H. Xin, B. Li, F. Meng, B. Hu, S. Wang, Y. Wang, and J. Li, 
``Quantitative proteomic analysis and verification identify global protein profiling dynamics in pig during the estrous cycle,'' 
Frontiers in Veterinary Science, vol. 10, 2023. [Online]. Available: 
\href{https://www.frontiersin.org/journals/veterinary-science/articles/10.3389/fvets.2023.1247561}{https://www.frontiersin.org/journals/veterinary-science/articles/10.3389/fvets.2023.1247561}\\[0pt]
[13] W. Jung, I. Yoo, J. Han, M. Kim, S. Lee, Y. Cheon, M. Hong, B.-Y. Jeon, and H. Ka, 
``Expression of caspases in the pig endometrium throughout the estrous cycle and at the maternal-conceptus interface during pregnancy and regulation by steroid hormones and cytokines,'' 
Frontiers in Veterinary Science, vol. 8, 2021. [Online]. Available: 
\href{https://www.frontiersin.org/journals/veterinary-science/articles/10.3389/fvets.2021.641916}{https://www.frontiersin.org/journals/veterinary-science/articles/10.3389/fvets.2021.641916}\\[0pt]
[14] W. Zhao, Q. Li, Y. Ma, Z. Wang, B. Fan, X. Zhai, M. Hu, Q. Wang, M. Zhang, C. Zhang, et al., 
``Behaviors related to psychiatric disorders and pain perception in C57BL/6J mice during different phases of estrous cycle,'' 
Frontiers in Neuroscience, vol. 15, 2021. [Online]. Available: 
\href{https://www.frontiersin.org/journals/neuroscience/articles/10.3389/fnins.2021.650793}{https://www.frontiersin.org/journals/neuroscience/articles/10.3389/fnins.2021.650793}\\[0pt]
[15] M. Milad, S. Igoe, K. Lebron-Milad, and J. Novales, 
``Estrous cycle phase and gonadal hormones influence conditioned fear extinction,'' 
Neuroscience, vol. 164, no. 3, pp. 887--895, 2009. [Online]. Available: 
\href{https://www.sciencedirect.com/science/article/pii/S0306452209014869}{https://www.sciencedirect.com/science/article/pii/S0306452209014869}\\[0pt]
[16] K. Lebron-Milad and M. Milad, 
``Sex differences, gonadal hormones and the fear extinction network: Implications for anxiety disorders,'' 
Biological Mood \& Anxiety Disorders, vol. 2, pp. 1--12, Jan. 2012.\\[0pt]
[17] B. Zhang, Y. Han, M. Cheng, L. Yan, K. Gao, D. Zhou, A. Wang, P. Lin, and Y. Jin, 
``Metabolomic effects of intrauterine meloxicam perfusion on histotroph in dairy heifers during diestrus,'' 
Frontiers in Veterinary Science, vol. 12, 2025. [Online]. Available: 
\href{https://www.frontiersin.org/journals/veterinary-science/articles/10.3389/fvets.2025.1528530}{https://www.frontiersin.org/journals/veterinary-science/articles/10.3389/fvets.2025.1528530}\\[0pt]
[18] A. Abulaiti, M. Nawaz, Z. Naseer, Z. Ahmed, W. Liu, M. Abdelrahman, A. Shaukat, A. Sabek, X. Pang, and S. Wang, 
``Administration of melatonin prior to modified synchronization protocol improves the productive and reproductive efficiency of Chinese crossbred buffaloes in low breeding season,'' 
Frontiers in Veterinary Science, vol. 10, 2023. [Online]. Available: 
\href{https://www.frontiersin.org/journals/veterinary-science/articles/10.3389/fvets.2023.1118604}{https://www.frontiersin.org/journals/veterinary-science/articles/10.3389/fvets.2023.1118604}\\[0pt]
[19] K. H. Kulkarni, Z. Yang, T. Niu, and M. Hu, 
``Effects of estrogen and estrus cycle on pharmacokinetics, absorption, and disposition of genistein in female Sprague–Dawley rats,'' 
Journal of Agricultural and Food Chemistry, vol. 60, no. 32, pp. 7949--7956, 2012.\\[0pt]
[20] T. A. Lovick and H. Zangrossi Jr, 
``Effect of estrous cycle on behavior of females in rodent tests of anxiety,'' 
Frontiers in Psychiatry, vol. 12, p. 711065, 2021.\\[0pt]
[21] S. L. Byers, M. V. Wiles, S. L. Dunn, and R. A. Taft, 
``Mouse estrous cycle identification tool and images,'' 
PLOS ONE, vol. 7, no. 4, pp. 1--5, Apr. 2012. [Online]. Available: 
\href{https://doi.org/10.1371/journal.pone.0035538}{https://doi.org/10.1371/journal.pone.0035538}\\[0pt]
[22] J. M. Goldman, A. S. Murr, and R. L. Cooper, 
``The rodent estrous cycle: Characterization of vaginal cytology and its utility in toxicological studies,'' 
Birth Defects Research Part B: Developmental and Reproductive Toxicology, vol. 80, no. 2, pp. 84--97, 2007. [Online]. Available: 
\href{https://api.semanticscholar.org/CorpusID:1021849}{https://api.semanticscholar.org/CorpusID:1021849}\\[0pt]
[23] A. Gal, P.-C. Lin, A. M. Barger, A. L. MacNeill, and C. Ko, 
``Vaginal fold histology reduces the variability introduced by vaginal exfoliative cytology in the classification of mouse estrous cycle stages,'' 
Toxicologic Pathology, vol. 42, no. 8, pp. 1212--1220, Dec. 2014. [Online]. Available: 
\href{https://europepmc.org/articles/PMC4378616}{https://europepmc.org/articles/PMC4378616}\\[0pt]
[24] J. K. MacDonald, W. G. Pyle, C. J. Reitz, and S. E. Howlett, 
``Cardiac contraction, calcium transients, and myofilament calcium sensitivity fluctuate with the estrous cycle in young adult female mice,'' 
American Journal of Physiology - Heart and Circulatory Physiology, vol. 306, no. 7, pp. H938--H953, Apr. 2014. [Online]. Available: 
\href{https://doi.org/10.1152/ajpheart.00730.2013}{https://doi.org/10.1152/ajpheart.00730.2013}\\[0pt]
[25] M. C. Cora, L. Kooistra, and G. S. Travlos, 
``Vaginal cytology of the laboratory rat and mouse,'' 
Toxicologic Pathology, vol. 43, pp. 776--793, 2015. [Online]. Available: 
\href{https://api.semanticscholar.org/CorpusID:32314150}{https://api.semanticscholar.org/CorpusID:32314150}\\[0pt]
[26] S. Matsuda, D. Matsuzawa, D. Ishii, H. Tomizawa, C. Sutoh, and E. Shimizu, 
``Sex differences in fear extinction and involvements of extracellular signal-regulated kinase (ERK),'' 
Neurobiology of Learning and Memory, vol. 123, pp. 117--124, 2015. [Online]. Available: 
\href{https://api.semanticscholar.org/CorpusID:207262888}{https://api.semanticscholar.org/CorpusID:207262888}\\[0pt]
[27] C. Hubscher, D. Brooks, and J. Johnson, 
``A quantitative method for assessing stages of rat estrous cycle,'' 
Biotechnic \& Histochemistry: Official Publication of the Biological Stain Commission, vol. 80, pp. 79--87, Mar. 2005.\\[0pt]
[28] G. Li, Z. Yu, K. Yang, M. Lin, and C. P. Chen, 
``Exploring feature selection with limited labels: A comprehensive survey of semi-supervised and unsupervised approaches,'' 
IEEE Transactions on Knowledge and Data Engineering, 2024.\\[0pt]
[29] W. Chen, K. Yang, Z. Yu, Y. Shi, and C. P. Chen, 
``A survey on imbalanced learning: Latest research, applications and future directions,'' 
Artificial Intelligence Review, vol. 57, no. 6, p. 137, 2024.\\[0pt]
[30] M. H. Hesamian, W. Jia, X. He, and P. J. Kennedy, 
``Deep learning techniques for medical image segmentation: Achievements and challenges,'' 
Journal of Digital Imaging, vol. 32, pp. 582--596, 2019. [Online]. Available: 
\href{https://api.semanticscholar.org/CorpusID:169033851}{https://api.semanticscholar.org/CorpusID:169033851}\\[0pt]
[31] A. W. Salehi, S. Khan, G. Gupta, B. I. Alabduallah, A. Almjally, H. Alsolai, T. Siddiqui, and A. Mellit, 
``A study of CNN and transfer learning in medical imaging: Advantages, challenges, future scope,'' 
Sustainability, vol. 15, no. 7, 2023. [Online]. Available: 
\href{https://www.mdpi.com/2071-1050/15/7/5930}{https://www.mdpi.com/2071-1050/15/7/5930}\\[0pt]
[32] S. P. Singh, L. Wang, S. Gupta, H. Goli, P. Padmanabhan, and B. Gulyás, 
``3D deep learning on medical images: A review,'' 
Sensors, vol. 20, no. 18, 2020. [Online]. Available: 
\href{https://www.mdpi.com/1424-8220/20/18/5097}{https://www.mdpi.com/1424-8220/20/18/5097}\\[0pt]
[33] R. Ding, X. Zhou, D. Tan, Y. Su, C. Jiang, G. Yu, and C. Zheng, 
``A deep multi-branch attention model for histopathological breast cancer image classification,'' 
Complex \& Intelligent Systems, vol. 10, Mar. 2024.\\[0pt]
[34] H. You, L. Yu, S. Tian, and W. Cai, 
``A stereo spatial decoupling network for medical image classification,'' 
Complex \& Intelligent Systems, vol. 9, Apr. 2023.\\[0pt]
[35] H.-C. Shin, H. R. Roth, M. Gao, L. Lu, Z. Xu, I. Nogues, J. Yao, D. Mollura, and R. M. Summers, 
``Deep convolutional neural networks for computer-aided detection: CNN architectures, dataset characteristics and transfer learning,'' 
IEEE Transactions on Medical Imaging, vol. 35, no. 5, pp. 1285--1298, 2016.\\[0pt]
[36] K. Yang, Z. Yu, W. Chen, Z. Liang, and C. P. Chen, 
``Solving the imbalanced problem by metric learning and oversampling,'' 
IEEE Transactions on Knowledge and Data Engineering, 2024.\\[0pt]
[37] M. A. Haq, I. Khan, A. Ahmed, S. M. Eldin, A. Alshehri, and N. A. Ghamry, 
``DCNNBT: A novel deep convolution neural network-based brain tumor classification model,'' 
Fractals, vol. 31, no. 06, p. 2340102, 2023. [Online]. Available: 
\href{https://doi.org/10.1142/S0218348X23401023}{https://doi.org/10.1142/S0218348X23401023}\\[0pt]
[38] B. Liao, H. Zuo, Y. Yu, and Y. Li, 
``GraphMRINet: A few-shot brain tumor MRI image classification model based on Prewitt operator and graph isomorphic network,'' 
Complex \& Intelligent Systems, pp. 1--14, 2024.\\[0pt]
[39] T. C. I. of Health Research, 
``Health portfolio sex- and gender-based analysis plus policy: Advancing equity, diversity and inclusion,'' 
Website, 2023. [Online]. Available: 
\href{https://www.canada.ca/en/health-canada/corporate/transparency/heath-portfolio-sex-gender-based-analysis-policy.html}{https://www.canada.ca/en/health-canada/corporate/transparency/heath-portfolio-sex-gender-based-analysis-policy.html}\\[0pt]
[40] K. Sano, S. Matsuda, S. Tohyama, D. Komura, E. Shimizu, and C. Sutoh, 
``Deep learning-based classification of the mouse estrous cycle stages,'' 
Scientific Reports, vol. 10, Jul. 2020.\\[0pt]
[41] N. S. Wolcott, K. K. Sit, G. Raimondi, T. Hodges, R. M. Shansky, L. A. Galea, L. E. Ostroff, and M. J. Goard, 
``Automated classification of estrous stage in rodents using deep learning,'' 
Scientific Reports, vol. 12, no. 1, p. 17685, 2022.\\[0pt]
[42] B. Babaev, S. Goyal, T. Arora, A. Autry, and R. A. Ross, 
``A novel method for estrous cycle staging using supervised object detection,'' 
NPP—Digital Psychiatry and Neuroscience, vol. 3, no. 1, p. 3, 2025.\\[0pt]
[43] C. Paccola, C. Resende, T. Stumpp, S. Miraglia, and I. Cipriano, 
``The rat estrous cycle revisited: A quantitative and qualitative analysis,'' 
Animal Reproduction (AR), vol. 10, no. 4, pp. 677--683, 2018.\\[0pt]
[44] M. Tan and Q. V. Le, 
``EfficientNet: Rethinking model scaling for convolutional neural networks,'' 
CoRR, vol. abs/1905.11946, 2019. [Online]. Available: 
\href{http://arxiv.org/abs/1905.11946}{http://arxiv.org/abs/1905.11946}\\[0pt]
[45] C. Liu, Z. Wei, L. Zhou, and Y. Shao, 
``Multidimensional time series classification with multiple attention mechanism,'' 
Complex \& Intelligent Systems, vol. 11, Nov. 2024.\\[0pt]

\end{document}